\documentclass[showpacs, onecolumn,nofootinbib]{revtex4}
\usepackage{graphicx}
\usepackage{color}
\usepackage{amsmath}

\begin{document}

\title{Investigating the dynamical models of cosmology with recent observations and upcoming gravitational-wave data}

\author{Jie Zheng}
\affiliation{Gravitational Wave and Cosmology Laboratory, Department of Astronomy, Beijing Normal University, Beijing 100875, China}
\author{Yun Chen*}
\author{Tengpeng Xu}
\affiliation{Key Laboratory for Computational Astrophysics, National Astronomical Observatories, Chinese Academy of Sciences, Beijing 100101, China}
\affiliation{College of Astronomy and Space Sciences, University of Chinese Academy of Sciences, Beijing, 100049, China}
\author{Zong-Hong Zhu*}
\affiliation{Gravitational Wave and Cosmology Laboratory, Department of Astronomy, Beijing Normal University, Beijing 100875, China}
\affiliation{School of Physics and Technology, Wuhan University, Wuhan 430072, China}

\email{chenyun@bao.ac.cn, zhuzh@bnu.edu.cn }




\begin{abstract}

We explore and compare the capabilities of the recent observations of standard cosmological probes and the future observations of gravitational-wave (GW) standard sirens on constraining cosmological parameters. It is carried out in the frameworks of  two typical dynamical models of cosmology, i.e., the $\omega_0\omega_a$CDM model with $\omega(z) =
\omega_0 +\omega_a*z/(1+z)$, and the $\xi$-index model with $\rho_X\propto\rho_ma^{\xi}$, where $\omega(z)$ is the dark energy equation of state, and $\rho_X$ and $\rho_m$ are the energy densities of dark energy and matter, respectively. In the cosmological analysis, the employed data sets include the recent observations of the standard cosmological probes, i.e., Type Ia supernovae (SNe Ia), baryon acoustic oscillation (BAO) and cosmic microwave background (CMB), and also the mock GW standard siren sample with 1000 merging neutron star events anticipated from the third-generation detectors. In the scenarios of both $\omega_0\omega_a$CDM and $\xi$-index models, it turns out that the mock GW sample can reduce the uncertainty of the Hubble constant $H_0$ by about 50\% relative to that from 
the joint SNe+BAO+CMB sample; nevertheless, the SNe+BAO+CMB sample demonstrates better performance on limiting other parameters.
Furthermore,  the Bayesian evidence is applied to compare the dynamical models with the $\Lambda$CDM model. The Bayesian evidences computed from the SNe+BAO+CMB sample reveal that the $\Lambda$CDM model is the most supported one; moreover, the $\omega_0\omega_a$CDM model is more competitive than the $\xi$-index model.

\end{abstract}

\maketitle

\section{Introduction}
\label{intro}
The existence of an exotic form of energy with negative pressure, dubbed ``dark energy'', is one of the most widely involved mechanism to explain the accelerating universe \cite{Huterer_Shafer_2018}. The most popular dark energy models mainly include the $\Lambda$CDM model and the scalar-field dark energy model \cite{Peebles_Ratra_2003}. Moreover, the $\Lambda$CDM model is preferred by most observations, though a small number of observations display slight deviations \cite{Bull_et_al_2016,Bullock_and_Boylan-Kolchin_2017}. However, on the theoretical level the $\Lambda$CDM model is embarrassed by the well-known cosmological constant problems \cite{Weinberg_1989,Carroll_1992}, i.e., the  ``coincidence'' and ``fine-tuning'' problems. The ``coincidence problem'' states that why the present epoch is so special that the energy density of dark energy is in the same order of magnitude as that of the matter only at this period. Several possible approaches have been adopted to explain or mitigate the coincidence problem, mainly including the anthropic principle \cite{Weinberg_2000,Vilenkin_2001,Garriga_et_al_2000,Garriga_and_Vilenkin_2001}, the slow evolving and spatially homogeneous scalar field with the ``tracker'' properties \cite{Copeland_et_al_2006}, and the interaction between the dark energy and dark matter \cite{Amendola_2000,Caldera-Cabral_et_al_2009}. 

In this work, we choose to explore two dynamical models of cosmology with the recent observations of standard cosmological probes and also the mock GW standard siren data.
One is the popular $\omega_{0}\omega_{a}$CDM model, where the equation of state (EoS) of dark energy is expressed with the Chevallier-Polarski-Linder (CPL) parameterization, i.e., $\omega(z)=\omega_{0}+\omega_{a}\frac{z}{1+z}$ \cite{CPL,CPL2}.  It is natural to consider a time-varying EoS $\omega(z)$ rather than a constant one like that in the $\Lambda$CDM scenario. Before the CPL parameterization proposed, there is another form of parameterized EoS, i.e., $\omega(z)=\omega_{0}+\omega_{a} z$. 
Nevertheless, the linear $\omega(z)$ has been gradually abandoned, mainly due to its increasing divergence at high redshift. The CPL parameterization can not only avoid divergence  but also reconstruct many scalar field equations of state with high accuracy\cite{CPL2}.
Another dynamical model considered here is a  phenomenological one which parameterizes the ratio of the energy densities of dark energy and matter (which contains the baryonic and dark matter) as $\rho_{X} \propto \rho_{m} a^{\xi}$ \cite{Dalal_et_al_2001}. It corresponds to two special cases, i.e., $\rho_X \propto \rho_m a^3$ with $\xi = 3$ for the $\Lambda$CDM model and  $\rho_X \propto \rho_m a^0$ with $\xi = 0$ for the self-similar solution without the coincidence problem. For simplicity, we name this phenomenological model ``$\xi$-index model''. In the $\xi$-index model, $\xi$ indicates the severity of the coincidence problem and how strongly $\rho_{X}/\rho_{m}$ varies over redshift. 

Benefiting from the multi-messenger era, we use not only the recent observations of standard cosmological probes but also the mock GW sample expected from the upcoming experiments in the cosmological analysis. 
Here, we employ the recent observations of standard cosmological probes including the SNe Ia data from the latest ``Pantheon" sample \cite{Scolnic_et_al_2018}, the BAO data from the measurements of 6dFGS survey\cite{Beutler2011}, SDSS DR7 MGS\cite{Ross2015the}, and BOSS DR12\cite{Alam2017} and the CMB power spectrum data from the Planck 2018 final analysis \cite{Aghanim_et_al_2018}. 
Owing to the feasibility of using GW sources as standard sirens \cite{Schutz_1986}, a certain number of works have tried to study the potential of GW events on constraining cosmological parameters \cite{2017ligo,Caironggen,zhang2019gw,zhangjingfei}. Due to the lack of real observational data of GW events at present, we choose to use the mock GW data generated by Du and Xu (2022) \cite{duminghui_gw} in this work. 

The rest of the paper is organized as follows. In Section 2, we briefly introduce the dynamical models under consideration, i.e., the $\omega_{0}\omega_{a}$CDM model and the $\xi$-index model. The data sets adopted in this work are described in Section 3. The results from observational constraints and the corresponding statistical analysis are displayed in Section 4. In the last section, we summarize the main conclusions.

\section{Two typical dynamical models of cosmology}
\label{sec:1}
In this section, we briefly introduce the two typical dynamical models investigated in this work, including the $\omega_{0}\omega_{a}$CDM model which has been widely studies in previous works \cite{lixiaolei,liubin,zhangjingfei,zhaozewei2020,qijingzhao} and the $\xi$-index model which also has been studied with several different cosmological probes \cite{Pavon_et_al_2004,Guo_et_al_2007,Chen_et_al_2010,Cao_et_al_2011,Zhang_et_al_2014}.

\subsection{$\omega_{0}\omega_{a}$CDM model}

In the framework of $\omega_{0}\omega_{a}$CDM model, the dark energy component is described with a time-varying EoS, 
\begin{equation}
    \omega(z)=\omega_{0}+\omega_{a}\frac{z}{1+z},
\end{equation}
which is the so-called CPL parameterization.
The $\omega_{0}\omega_{a}$CDM model as one of the extensions of the $\Lambda$CDM model is widely used in the cosmological analysis,  which can reduce to the $\Lambda$CDM model with $(\omega_{0},\omega_a)=(-1,0)$.
In the flat Friedmann–Lemaitre–Robertson–Walker (FLRW) universe, the Friedmann equation of $\omega_{0}\omega_{a}$CDM model is expressed as,
\begin{eqnarray}
\label{eq:E_CPL}
E^{2}\left(z ; \mathbf{p}\right)&=\Omega_{m,0}(1+z)^{3}+\left(1-\Omega_{m,0}\right) \times \\
&(1+z)^{3\left(1+w_{0}+w_{a}\right)} \exp \left(\frac{-3 w_{a} z}{1+z}\right), \nonumber
\end{eqnarray}
where $E(z ; \mathbf{p})\equiv H(z)/H_0$ is the dimensionless Hubble parameter, and the model parameters are $\mathbf{p}=(\Omega_{m,0},\omega_{0},\omega_{a},H_{0})$.
The CPL form has been widely applied since it was proposed, which mainly owes to its good performance at both low and high reshifts \cite{CPL,CPL2}.

\subsection{$\xi$-index model}
A phenomenological model with an assumption for the ratio of the dark energy and matter densities is proposed in \cite{Dalal_et_al_2001}, i.e.,
\begin{equation}
\rho_{X} \propto \rho_{m} a^{\xi}, \qquad or \qquad \Omega_{X} \propto \Omega_{m} a^{\xi},
\end{equation}
where $\Omega_{X}$ and $\Omega_{m}$ are the fractions of the energy density of the universe contributed from dark energy and matter, respectively. In this model, the most crucial parameter ${\xi}$ can well reveal the severity of the coincidence problem, so we have named this model ``$\xi$-index model''.
This parametrization originates from two special cases, i.e., when $\xi=3$ it recovers to the $\Lambda$CDM model and when $\xi=0$ it corresponds to the self-similar solution without the coincidence problem.
 Furthermore,
any solution with a scaling parameter $0<{\xi}<3$ makes the coincidence problem less severe \cite{Pavon_et_al_2004}.
Considering a flat FLRW universe with $\Omega_{X}+\Omega_{m}=1$, one can obtain
\begin{equation}
\Omega_{X} =\frac{\Omega_{X,0} a^{\xi}}{1-\Omega_{X,0}\left(1-a^{\xi}\right)},
\end{equation}
where $\Omega_{X,0}=\Omega_{X}(z=0)$. According to the energy conservation equation, one has
\begin{equation}
 \frac{d \rho_{\mathrm{tot}}}{d a}+\frac{3}{a}\left(1+\omega_{X} \Omega_{X}\right) \rho_{\mathrm{tot}}=0,
\label{eq:equation2}
\end{equation}
where $\rho_{\mathrm{tot}}=\rho_{m}+\rho_{X}$ is the total energy density, $\omega_{X}$ specifies the EoS of the dark energy. Meanwhile, the Eq.(\ref{eq:equation2}) can be rewritten as
\begin{equation}
\frac{d \rho_{m}}{d a}+\frac{3}{a} \rho_{m} = -\left[\frac{d \rho_{X}}{da}+\frac{3}{a}\left(1+\omega_{X}\right) \rho_{X}\right] = Q,
\end{equation}
where $Q = -(\xi+3\omega_X)\rho_m \kappa a^{\xi-1}/(1+\kappa a^{\xi})$ and $\kappa = \rho_X/(\rho_m a^{\xi})$, and the interaction term
$Q = 0\;(\neq 0)$ denotes the cosmology without (with) interaction between dark energy and matter. The cases of $Q = 0\;(\neq 0)$ can be equal to $\xi + 3\omega_{X}= 0\;(\neq 0)$. Furthermore, $\xi + 3\omega_{X} > 0$, corresponding to $Q < 0$,  implies that the energy is transferred from matter to dark energy; oppositely, $\xi + 3\omega_{X} < 0$, corresponding to $Q > 0$, denotes that the energy is transferred from dark energy to dark matter.

Based on Eq.(\ref{eq:equation2}), one can work out
\begin{equation}
E^{2}(z)=\frac{\rho_{\mathrm{tot}}}{\rho_{0}}=\exp \left(\int_{a}^{1} \frac{d a}{a} 3\left(1+\omega_{X} \Omega_{X}\right)\right).
\label{eq:equation4}
\end{equation}
When taking $\omega_{X}$ and $\xi$ as constants, one can solve Eq.(\ref{eq:equation4}) and get
\begin{equation}
\label{eq:equation5}
E^{2}(z;\textbf{p})=a^{-3}\left[1-(1-\Omega_{m,0}) 
\times(1-a^{\xi})\right]^{-3 \omega_{X} / \xi},
\end{equation}
where the model parameters are $\textbf{p} \equiv \left(\Omega_{m,0}, \omega_{X}, \xi, H_{0}\right)$.

\section{Data sample}

\begin{figure*}
\centering 
\includegraphics[scale=0.6, angle=0] {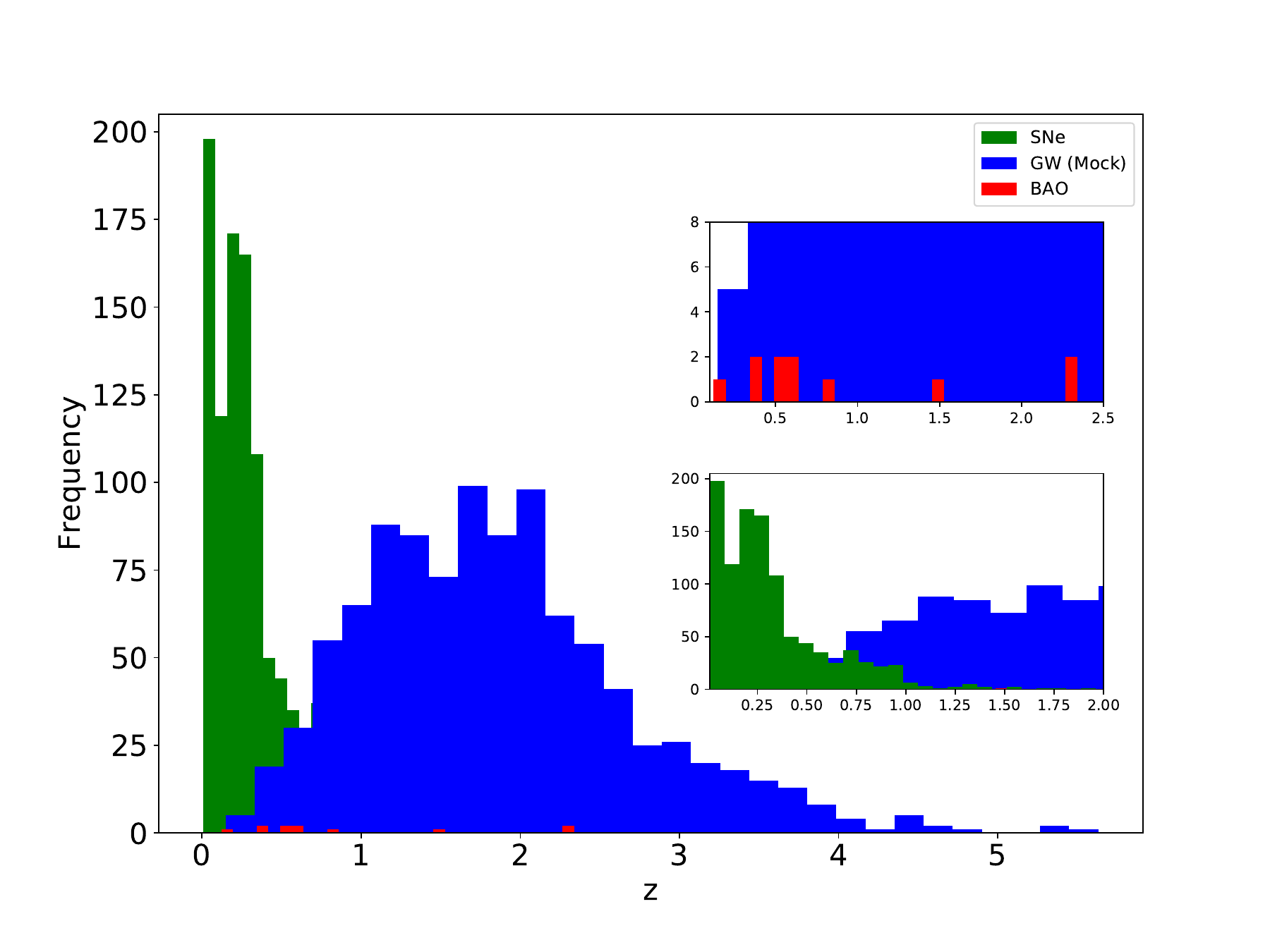}
\caption {The redshift distribution of the standard cosmological observations and the mocked GW data. Green, blue and red histograms represent the redshift distribution of SNe, the mocked GW and BAO, respectively.}
\label{fig:zf}
\end{figure*}

The recent observations of the standard cosmological probes and the mock data of the third-generation GW detection are both employed in our cosmological analysis.
The adopted data of the standard cosmological probes include the Pantheon SNe Ia sample, and the BAO data from the 6dFGS survey, the SDSS DR7 MGS, and the BOSS DR12 measurements, along with the CMB power spectrum data from the final Planck 2018 results. 
The redshift distributions of the mock GW data and the SNe and BAO data are displayed in Fig.\ref{fig:zf}.

\subsection{SNe Ia data set}
The SNe Ia as standard candles have been proved to be a kind of sensitive probe of cosmology (see, e.g. \cite{Branch_and_Miller_1993,Riess_Press_and Kirshner_1996,Filippenko_2005}).
The population of confirmed SNe Ia has a dramatic increase over the last two decades, in the mean time, the techniques for measuring the light curve parameters are also continually being improved to reduce the systematic uncertainties. At present,
the most popular techniques mainly include the SALT/SALT2 \cite{Guy_et_al_2005,Guy_et_al_2007} and SiFTO \cite{Conley_et_al_2008} models, which are two popular techniques at present and fit the light curves of SNe Ia by using the spectral template.

The SNe Ia sample adopted in this work is the Pantheon sample \cite{Scolnic_et_al_2018}, which consists of 1048 SNe Ia (0.01 $\le z \le$ 2.3) combined from Pan-STARRS1(PS1) Medium Deep Survey, SDSS, SNLS, various low-z and HST samples.
In the Pantheon sample, the distances for each of these SNe Ia are determined after fitting their light-curves with the most up-to-date published version of SALT2 \cite{Betoule_et_al_2014}, then applying the BEAMS with Bias Corrections (BBC) method \cite{Kessler_and_Scolnic_2017} to determine the nuisance parameters and adding the distance bias corrections. The uniform analysis procedure conducted on the SNe Ia of Pantheon sample has significantly reduced the systematic uncertainties related to photometric calibration.

The observable given in the Pantheon sample can be deemed as a correction to the apparent magnitude (see Table A17 of \cite{Scolnic_et_al_2018}), i.e.,
\begin{eqnarray}
Y^{obs} &=& m_B+K \nonumber\\
        &=& \mu+M,
\label{eq:Y_obs}
\end{eqnarray}
where $\mu$ is the distance modulus, $m_B$ is the apparent B-band magnitude, $M$ is the absolute B-band magnitude of a fiducial SN Ia, and the correction term $K = \alpha x_1-\beta c+\Delta_M+\Delta_B$ includes the corrections related to four different sources (see \cite{Scolnic_et_al_2018} for more details). The corresponding theoretical (predicted) value is
\begin{eqnarray}
Y^{th}&=& 5\log(d_L)+25 +M \nonumber\\
&=&5\log[(1+z)D(z)]+ Y_0,
\label{eq:Y_th}
\end{eqnarray}
where the constant term $Y_0$ is written as $Y_0 = M+5log(\frac{cH_0^{-1}}{Mpc})+25$, and the luminosity distance $d_L$ and the normalized comoving distance $D(z)$ are related with each other through the following formula, i.e.,
\begin{equation}
\label{eq:dl}
d_L(z) = \frac{c(1 + z)}{H_0}D(z),
\end{equation}
where $c$ is the velocity of light.
In a flat universe,  $D(z)$ can be expressed as
\begin{equation}
D(z) = \int_0^z\frac{d\tilde{z}}{E(\tilde{z})},
\label{eq:D_z}
\end{equation}
where $E(z)$ can be worked out with Eq.(\ref{eq:E_CPL}) and Eq. (\ref{eq:equation4}) for the model under consideration.

The chi-square statistic for the Pantheon sample can be constructed as
\begin{equation}
\label{eq:chi2SNe}
\chi^2_{\textrm{SNe}}={\Delta \overrightarrow{Y}}^T\cdot\textbf{C}^{-1}\cdot{\Delta \overrightarrow{Y}},
\end{equation}
where the residual vector for the SNe Ia data in the Pantheon sample is $\Delta \overrightarrow{Y}_i = [Y^{obs}_i-Y^{th}(z_i; Y_0,\textbf{p})]$. The covariance matrix $\textbf{C}$ of the sample includes the contributions from both the statistical and systematic errors. The nuisance parameter, i.e., the constant term $Y_0$ is marginalized over with the analytical methodology presented in \cite{Giostri_et_al_2012}.

\subsection{BAO data set} 

The BAO data extracted from galaxy redshift surveys are also a kind of  powerful cosmological
probe \cite{eisenstein1998baryonic,eisenstein2005detection}.
The BAO data set used here is 
a combination of measurements from the
6dFGS at $z_{\rm{eff}}=0.106$ with the corresponding measurement of $r_{s}/D_{V}$ described in \cite{Beutler2011}, the SDSS DR7 Main
Galaxy Sample (MGS) at $z_{\rm{eff}}=0.15$ presented in \cite{Ross2015the} with the corresponding measurement of $D_{V}(r_{s,fid/r_{s}})$, and the BOSS DR12 at $z_{\rm{eff}} = (0.38,0.51,0.61)$ with the corresponding measurements of $D_{M}(r_{s,fid}/r_{s})$ and $H(z)(r_{s}/r_{s,fid})$ \cite{Alam2017}.

The observable quantities used in the measurements are expressed in terms of the transverse co-moving distance $D_M(z)$, the volume-averaged angular diameter distance $D_V(z)$, the Hubble rate $H(z)\equiv H_0E(z)$,  the sound horizon at the drag epoch $r_s$, and its fiducial value $r_{\rm{s,fid}}$. 
Following \cite{Ryan_Chen_Ratra_2019}, we use the fitting formula of \cite{eisenstein1998baryonic} to compute $r_s$, and $r_{\rm{s,fid}}$ is computed with the fiducial cosmology adopted in the paper in which the measurement is reported.
In a flat universe, the transverse co-moving distance $D_M(z)$ equals to the line-of-sight comoving distance $D_{C}(z)$, which is expressed as,
\begin{equation}
D_{C}(z) \equiv \frac{c}{H_{0}} D(z),
\label{eq:equationDC}
\end{equation}
and $c$ is the speed of light.
The volume-averaged angular diameter distance is \begin{equation}
D_{V}(z)=\left[\frac{c z}{H_{0}} \frac{D_{M}^{2}(z)}{E(z)}\right]^{1 / 3}.
\label{eq:equationDVz}
\end{equation}
We employ the BAO data set in the analysis with the chi-square:
\begin{equation}
\chi_{\mathrm{BAO}}^{2}(p)=\left[\vec{A}_{\mathrm{th}}(p)-\vec{A}_{\mathrm{obs}}\right]^{T} C^{-1}\left[\vec{A}_{\mathrm{th}}(p)-\vec{A}_{\mathrm{obs}}\right],
\label{eq:chi2_BAO}
\end{equation}
where $C^{-1}$ is the inverse of the covariance matrix. The BOSS DR12 measurements are correlated, and the corresponding covariance matrix is present in Eq.(20) of \cite{Ryan_Chen_Ratra_2019}, which is also available from SDSS website \footnote{\url{https://sdss3.org/science/boss\_publications.php}}.

\subsection{CMB data set} 
Observations of the CMB spectra provide another kind of independent test of the existence of dark energy.
It is remarkable that the CMB power spectra from the WMAP \cite{Hinshaw2013WMAP} and Planck projects \cite{Aghanim_et_al_2018} have provided strong constraints on cosmological parameters. Here, we use the combination of temperature and polarization CMB power spectra from the Planck 2018 release \cite{Aghanim_et_al_2018}, including the likelihoods at multipoles $\ell=2-2508$ in TT, $\ell=2-1996$ in EE, and $\ell=30-1996$ in TE. 
In practice, different algorithms have been used to estimate the CMB power spectrum, such as Commander\cite{Planck_2018_A4,Planck_2018_A5}, SimAll\cite{Planck_2018_A5} and Pilk\cite{{Aghanim_et_al_2018}}.
The ``Commander'' component-separation algorithm is used to estimate the power spectrum over the range  $\ell=2-29$ in TT. The ``SimAll'' approach is used to estimate the power spectrum over the range   $\ell=2-29$ in EE. 
The ``Pilk'' cross-half-mission likelihood \cite{Planck_2018_A5} is used to compute the CMB high-$\ell$ part for TT,TE,EE over the range $30 \leq \ell \leq  2508$ in TT and over the range $30 \leq \ell \leq 1996$ in TE and EE \footnote{ \url{https://wiki.cosmos.esa.int/planckpla/index.php}}. Hereafter, $\mathcal{L}_{Planck}$ denotes the likelihood of the Planck data described above.

\subsection{Simulated sample of GW events from third-generation detectors}
It is expected that the second-generation (2G) GW detectors will detect more and more nearby GW events routinely. Besides, the third-generation (3G) GW detectors will bring more hopes and opportunities in the cosmological applications of the GW standard sirens, since they have the potential to detect a large number of GW events at cosmological redshifts. 

Among the different types of GW sources, the GWs from binary neutron stars (BNS) are supposed to be most prospective ones in the cosmological applications, since they own more easily detectable electromagnetic (EM) counterparts, e.g., the coincident short gamma-ray burst (SGRB) events. As it is discussed in \cite{Belgacem_et_al_2019}, the expectation of 1000 events of  BNS GWs as standard sirens along with EM counterparts seems to be reasonable and promising with the observations of the 3G GW detectors and the future THESEUS mission. 

As to the BNS-SGRB GW events, the degeneracy between luminosity distance and inclination angle of the orbit is proved to be a chief factor which significantly affects the measurement of luminosity distance and further affects the cosmological implementation. However, most previous studies generated the mock data with the customary practice, i.e., neglecting the distance–inclination correlation and account for the influence of correlation by multiplying a factor of 2 \cite{Li_2015,Zhang_et_al_2020,Zhang_et_al_2019,Du_et_al_2019,Cai_Yang_2017,Zhao_et_al_2011}. Concerning this issue, Du and Xu (2022) \cite{duminghui_gw} has 
adopted a Gaussian prior on the cosine of inclination angle, where it is assumed that the viewing angle of SGRB is identical to the inclination angle, that is based on the theoretical expectation \cite{Nissanke2010,fanxilong2017}. In addition, the BNS GW events are just detectable in the sensitive bands of the 3G ground-based detectors\footnote{While the space detector LISA is good at detecting events of coalescing supermassive black hole binaries at large redshift.}, such as the Einstein Telescope (ET) in Europe, the Cosmic Explorer (CE) in the US, and an assumed CE-like detector in Australia. Following Du and Xu (2022) \cite{duminghui_gw}, we use ET + CE to denote the network of ET and two CE-like detectors. Du and Xu (2022) \cite{duminghui_gw} has generated four mock catalogs with 1000 BNS-SGRB GW events from 3G ground-based detectors with redshift ranging $0.148 \leq z \leq 5.631$, where the four catalogs originate from four different configurations of the detectors and the priors on the cosine of inclination angle (i.e., $ \nu = \cos \iota$). Here we choose to use the mock catalog corresponding to the configuration of ET+CE with the prior of $\nu = 1\pm 0.003$, which is the most optimistic one among the four configurations considered in \cite{duminghui_gw}.
Then, we can construct the chi-square statistic for the  mock sample of 1000 BNS-SGRB GW events as  
\begin{equation}
\chi_{GW}^{2}(p)=\frac{(d_{L,i}^{mock}-d_{L,i}^{th}(p))^{2}}{\sigma_{d_{L}}^{2}},
\end{equation}
where $d_{L}$ can be calculate with Eq.(\ref{eq:dl}).

\section{Analysis and discussion}
\subsection{Observational constraints}
\begin{table*}
\caption{\label{tab:parameters}
The mean values with 68\% confidence limits for the parameters of interest constrained from the  SNe+BAO+CMB sample and the mock GW sample, respectively. The scenarios of $\omega_{0}\omega_{a}$CDM model and  $\xi$-index model are both considered.}
\centering
\begin{tabular}{ccccccc}
\hline
\hline

Model & Data set &$\Omega_{m,0}$ & $\omega_{0}$ & $\omega_{a}$ & $H_{0}$ \\
\hline
$\omega_{0}\omega_{a}$CDM & SNe+BAO+CMB &$0.31 \pm 0.01$ &$-0.98^{+0.09}_{-0.06}$ & $-0.15^{+0.19}_{-0.35}$ & $68.25^{+0.85}_{-0.80}$\\
& GW (Mock) & $0.32^{+0.02}_{-0.01}$ & $-1.11^{+0.98}_{-0.65}$ & $0.04^{+0.61}_{-0.93}$ & $68.06^{+0.54}_{-0.45}$ \\ 

\hline
Model & Data set & $\Omega_{m,0}$ & $\omega_{X}$ & $\xi$ & $H_{0}$ \\
\hline
$\xi$-index & SNe+BAO+CMB  & $0.33 \pm0.01$ & $-1.18^{+0.06}_{-0.04}$ & $3.44^{+0.14}_{-0.19}$ & $66.43^{+0.83}_{-1.00}$ \\
& GW (Mock) & $0.39^{+0.13}_{-0.20}$ & $-1.24^{+0.38}_{-0.41}$ & $3.05^{+0.77}_{-0.45}$ & $68.23^{+0.56}_{-0.51}$ \\  

\hline
\hline
\end{tabular}
\end{table*}

\begin{table}
\caption{\label{tab:accuracy}
Constraint precisions of the parameters obtained from the SNe+BAO+CMB sample and the mock GW sample, respectively. The relative precision $\varepsilon(p)$ is defined as $\varepsilon(p)=\sigma(p)/p$.}
\centering
\begin{tabular}{ccccc}
\hline
\hline

$\omega_{0}\omega_{a}$CDM & $\varepsilon(\Omega_{m,0})$ & $\varepsilon(\omega_{0})$ & $\varepsilon(\omega_{a})$ & $\varepsilon(H_{0})$  \\
\hline
SNe+BAO+CMB  &0.032& 0.078 & 1.877 & 0.012  \\
GW (Mock) & 0.049 & 0.749 & 19.661 & 0.007 \\ 

\hline
$\xi$-index & $\varepsilon(\Omega_{m,0})$ & $\varepsilon(\omega_{X})$ & $\varepsilon(\xi)$ & $\varepsilon(H_{0})$ \\
\hline
SNe+BAO+CMB & 0.030 & 0.043 & 0.049 & 0.014  \\
GW (Mock) & 0.432 & 0.319 & 0.207 & 0.008  \\  

\hline
\hline
\end{tabular}
\end{table}

\begin{figure*}
\centering 
\includegraphics[scale=0.65, angle=0] {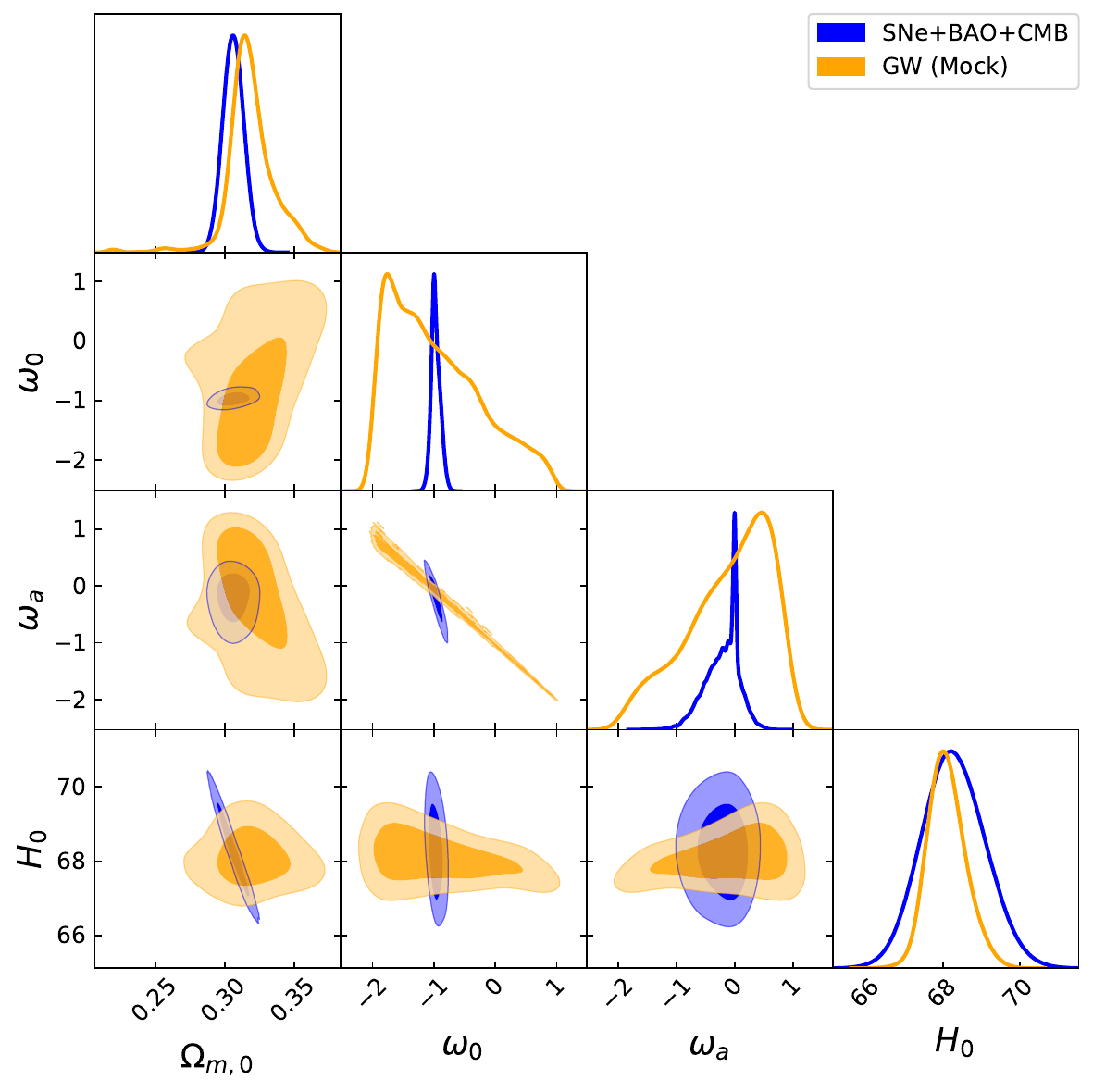}
\caption {The 1D and 2D probability distributions of parameters of interest in the scenario of $\omega_{0}\omega_{a}$CDM model constrained from the SNe+BAO+CMB sample (blue lines) and the mock GW sample (yellow lines), respectively. The contours correspond to 68\% and 95\% CLs.}
\label{fig:CPL}
\end{figure*}

\begin{figure*}
\centering 
\includegraphics[scale=0.65, angle=0] {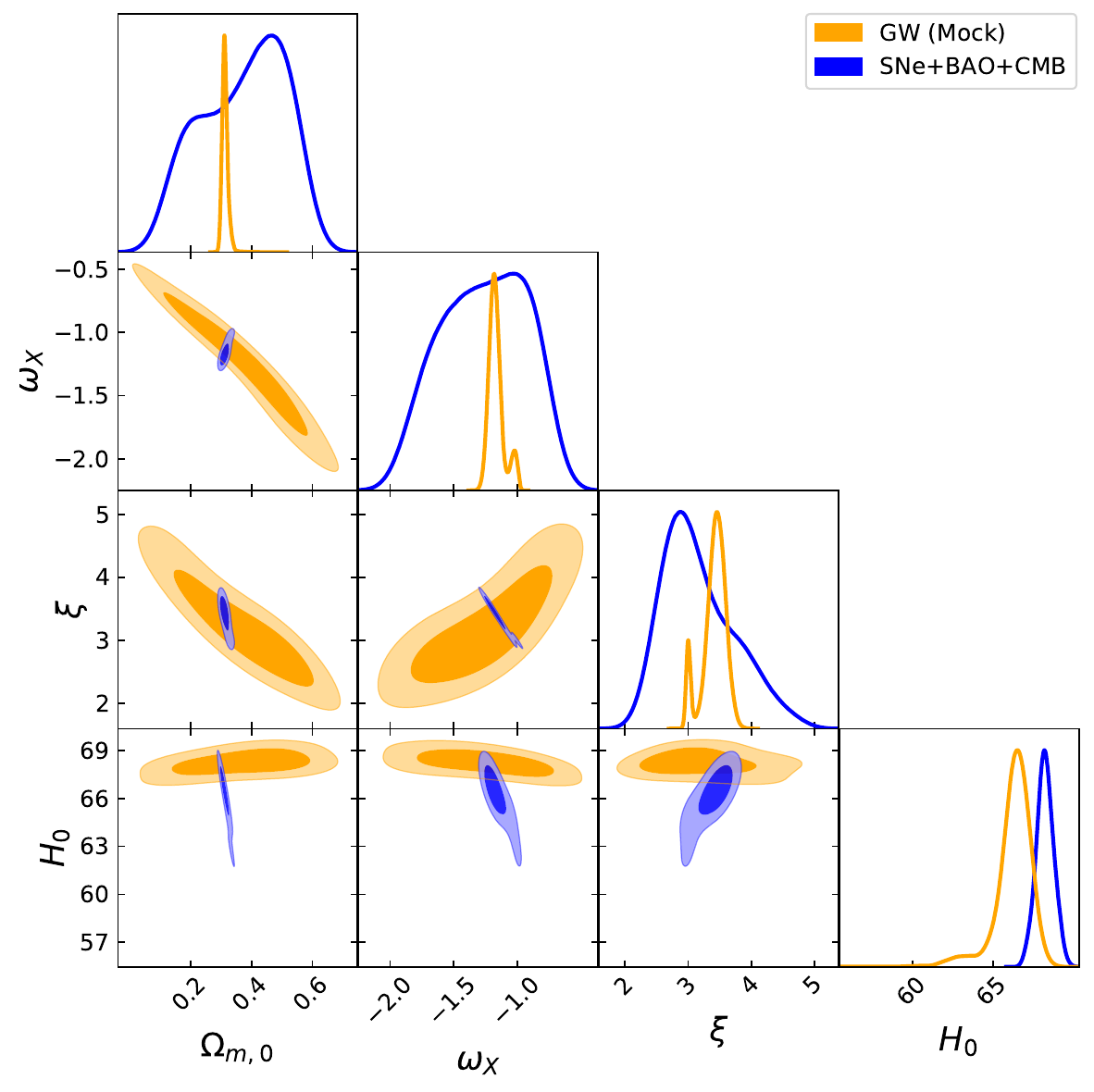}
\caption {The 1D and 2D probability distributions of parameters of interest in the scenario of $\omega_{0}\omega_{a}$CDM model constrained from the SNe+BAO+CMB sample (blue lines) and the mock GW sample (yellow lines), respectively. The contours correspond to 68\% and 95\% CLs.}
\label{fig:pmodel}
\end{figure*}

In our analysis, the total likelihood for parameters is
\begin{equation}
\label{eq:LH_total}
\mathcal{L}(\mathbf{p})=\prod \mathcal{L}_{i},
\end{equation}
where $\mathcal{L}_{i}$ means the likelihood of each data set, and $\mathcal{L}_{i}\propto e^{\chi_i^2/2}$ for the SNe, BAO and mock GW data sets.
In the case of using the combination of SNe Ia, BAO and CMB data sets, it takes
\begin{equation}
\mathcal{L}(\mathbf{p})=\mathcal{L}_{SNe} \mathcal{L}_{BAO} \mathcal{L}_{Planck}.
\end{equation}
 In the case of using the mock GW sample alone, it could be
\begin{equation}
\mathcal{L}(\mathbf{p})=\mathcal{L}_{GW}.
\end{equation}
We derive the posterior probability distributions of parameters with Markov Chain Monte Carlo (MCMC) exploration using the May 2020 version of CosmoMC \cite{Lewis_et_al_2002}.

The one-dimensional (1D) probability distributions and two-dimensional (2D) contours for the cosmological parameters of interest are shown in Fig.\ref{fig:CPL} for the $\omega_0\omega_a$CDM model and in Fig.\ref{fig:pmodel} for the $\xi$-index model. From Figs.\ref{fig:CPL} and \ref{fig:pmodel}, one can discover that the parameter degeneracy directions of the joint SNe+BAO+CMB sample are visibly different from those of the mock GW sample. Thus, the combination of them should be able to break the parameter degeneracies effectively, that will be applicable in future when more GW events are detected.
In the framework of $\omega_0\omega_a$CDM model, Fig. \ref{fig:CPL} shows that the joint SNe+BAO+CMB sample is more powerful on constraining the parameters $\Omega_{m,0}$, $\omega_{0}$ and $\omega_a$, while the mock GW sample is more sensitive to the parameter $H_0$.  In the framework of $\xi$-index model, Fig. \ref{fig:pmodel} reveals that the mock GW sample is more powerful on constraining $H_0$, while the joint SNe+BAO+CMB sample is more sensitive to other parameters, i.e., $\Omega_{m,0}$, $\omega_X$ and $\xi$. According to the above analyses, one can realize that the key advantage of the GW sample is its outstanding constraining capability on the parameter $H_0$.

The mean values of the parameters of interest together with their 68\% confidence limits are presented in Table \ref{tab:parameters} for both the $\omega_0\omega_a$CDM and $\xi$-index models.
Moreover, Table \ref{tab:accuracy} displays the relative error of each parameter defined as $\varepsilon(p)=\sigma(p)/p$, where $p$ denotes the mean value of the corresponding parameter, and the error $\sigma(p)$ is obtained with $\sigma(p) = \sqrt{(\sigma_{upper}^2 + \sigma_{lower}^2 )/2}$. 
In the scenario of $\omega_0\omega_a$CDM model, the constraint precision of $\Omega_{m,0}$ from the joint SNe+BAO+CMB sample is a little higher than that from the mock GW sample; however, the precisions of $\omega_0$ and $\omega_a$ from the SNe+BAO+CMB sample are much higher than those from the mock GW sample, where the former are around 10 times more precise than the latter. While the relative error of $H_0$ from the mock GW sample is about 50\% of that from the SNe+BAO+CMB sample. In the scenario of $\xi$-index model, the constraint precisions of $\Omega_{m,0}$, $\omega_X$ and $\xi$ from the joint SNe+BAO+CMB sample are all much higher than those from the mock GW sample; however, the precision of $H_0$ from the mock GW sample is about one time higher than that from the SNe+BAO+CMB sample. These results are consistent with those from Figs. \ref{fig:CPL} and \ref{fig:pmodel}.

\subsection{Model Selection Statistics}
\begin{table}
\centering
	\caption{The natural logarithm of the Bayesian evidences $\ln B_{i}$ and the Bayes factors $\ln B_{i,0}$ from the joint sample of SNe+BAO+CMB, where the subscript ``0'' denotes the $\Lambda$CDM model.}
	\begin{tabular}{ccc}
		\hline
		Model &  $\ln B_{i}$ & $\ln B_{i,0}$ \\
		\hline
		$\Lambda$CDM & -1940.80 & 0 \\
		$\omega_{0}\omega_{a}$CDM & -1942.82 & -2.02 \\
		$\xi$-index & -2022.22  & -81.42 \\
		\hline
	\end{tabular}

	\label{tab:table_selection}
\end{table}

In the framework of Bayes' theorem, the probability that the model $M_{i}$ is true can be estimated with 
\begin{equation}
P\left(M_{i} \mid D\right)=\frac{P\left(D \mid M_{i}\right) P\left(M_{i}\right)}{P(D)},
\end{equation}
where $P(M_{i} \mid D)$ is the posterior probability, $D$ denotes the observational data, $P(M_{i})$ is a prior probability in the model $M_{i}$, and  $P(D)$ is the normalization constant. In addition, $P(D \mid M_{i})$ is the so-called Bayesian evidence \cite{roberto2008,limitation2008}, which can be written as
\begin{equation}
P\left(D \mid M_{i}\right)=\int P\left(D \mid \bar{\theta}, M_{i}\right) P\left(\bar{\theta} \mid M_{i}\right) d \bar{\theta}, 
\end{equation}
where $P(D \mid \bar{\theta},M_{i})$ is the likelihood function under the model $M_i$, and $P(\bar{\theta} \mid M_{i})$ is the prior probability for parameter $\bar{\theta}$ under the model $M_i$. Hence, calculating the Bayesian evidence requires the evaluation of an integral over the entire likelihood function and the prior distributions of model parameters.
When comparing two models, e.g., $M_{i}$ versus $M_{j}$, the Bayes factor
\begin{equation}
B_{i j}=\frac{P\left(D \mid M_{i}\right)}{P\left(D \mid M_{j}\right)},
\end{equation}
which is defined as the ratio of the Bayesian evidences of two models can be employed as a judgment criterion, where the Bayes factor $B_{ij}>1$ (i.e., $\ln B_{ij} > 0$) means that the observational data prefer $M_i$ to $M_j$, and $B_{ij}<1$ implies that $M_j$ is preferred \cite{bayesfactor}. 
 
To compare the two phenomenological models under consideration with the $\Lambda$CDM model, we calculate the values of Bayesian evidence for each model, where the code $\textbf{MCEvidence}$ \cite{MCEvidence} which is a popular python package to compute the Bayesian evidence is adopted here, and the observational data correspond to the joint sample of SNe, BAO and CMB data. 
In Table \ref{tab:table_selection}, we present the natural logarithm of the Bayesian evidence for each model, $\ln B_i$, as well as the natural logarithm of the Bayes factor, $\ln B_{i0}$, where the subscript ``0'' denotes the $\Lambda$CDM model.
It turns out that the $\Lambda$CDM model is most supported by the joint sample, since $B_{1,0}$ and $B_{2,0}$ are both smaller than 1, where the subscripts ``1'' and ``2'' denote the scenarios of $\omega_{0}\omega_{a}$CDM model and $\xi$-index model, respectively. In addition, $B_{1,2} =B_1/B_2$ is bigger than 1, so the $\omega_{0}\omega_{a}$CDM model is more competitive than the $\xi$-index model. 

\section{Summary and conclusions} 
We have concentrated on two typical dynamical models of cosmology, i.e., the $\omega_0\omega_a$CDM model with $\omega(z)=\omega_0+\omega_a*(1-a)$, and the $\xi$-index model with $\rho_X\propto\rho_ma^{\xi}$.
To explore the capability of the future GW  standard siren data on constraining the cosmological parameters, the data sets employed in the cosmological analysis include the recent observations of the standard cosmological probes, i.e., SNe+BAO+CMB, and the mock BNS-SGRB GW sample anticipated from  the 3G GW experiments. The MCMC method is applied to obtain the posterior probability distributions of the model parameters.
Moreover, we also use the Bayesian evidence to compare the two dynamical models under consideration with the $\Lambda$CDM model.
The new points of this work mainly include:(i) the mock GW data used in this work are from Du and Xu (2022), which are generated based on more stringent assumptions to break the degeneracy between luminosity distance and inclination angle of the orbit, and thus are distinct from those used in other previous studies;(ii) in the previous studies, the forecasts from GW have been done in the frameworks of typical models (like $\Lambda$CDM and $\omega_0\omega_a$CDM models) rather than the $\xi$-index model considered in this work. 
The main conclusions can be summarized as follows:

(i) The GW standard siren sample demonstrates outstanding capability on limiting the parameter $H_0$. The mock sample with 1000 BNS-SGRB GW events expected from the 3G GW experiments can reduce the uncertainty of  $H_0$ by about 50\% relative to that from the recent SNe+BAO+CMB sample. While the joint SNe+BAO+CMB sample has better performance on constraining other parameters, i.e., $(\Omega_{m,0},\omega_0,\omega_a)$ in the $\omega_0\omega_a$CDM model and $(\Omega_{m,0},\omega_X,\xi)$ in the $\xi$-index model. Our results are consistent with those from Belgacem et al. (2019)\cite{Belgacem_et_al_2019}, where they found that the BNS-SGRB GW sample expected from the 3G GW experiments can reduce the uncertainty of  $H_0$ by more than 50\% relative to that from the SNe+BAO+CMB sample in the $\Lambda$CDM model. Comparing with \cite{Belgacem_et_al_2019}, the main improvements of this work include: i) the data sets of SNe, BAO and CMB are updated; ii)  when generating the mock data of GW events, the degeneracy
between luminosity distance and inclination angle of
the orbit was not considered in  \cite{Belgacem_et_al_2019}; while the mock GW sample used in this work was generated with the consideration of this degeneracy \cite{duminghui_gw}.  

(ii) In the scenarios of $\omega_0\omega_a$CDM and $\xi$-index models, the 2D contours for the cosmological parameters display that the parameter degeneracy directions of the mock GW sample are significantly different from those of the joint SNe+BAO+CMB sample. Thus,  it's anticipated that a considerable amount of GW events in future will help to break the parameter degeneracies existing in the standard cosmological probes. These tendencies have also emerged in the scenario of $\Lambda$CDM model as discussed in \cite{Belgacem_et_al_2019}.

 (iii) In the scenario of $\omega_0\omega_a$CDM model, the constraints from the SNe+BAO+CMB data set turn out to be $\omega_0 = -0.98^{+0.09}_{-0.06}$ and $\omega_a = -0.15^{+0.19}_{-0.35}$ at 68\% confidence level (CL), that is consistent with the $\Lambda$CDM scenario of $(\omega_0,\omega_a)=(-1,0)$ at 68\% CL.

(iv) In the scenario of $\xi$-index model, the constrains from the SNe+BAO+CMB data set display that $\omega_X = -1.18^{+0.06}_{-0.04}$ and $\xi = 3.44^{+0.14}_{-0.19}$ at 68\% CL, where the $\Lambda$CDM model, corresponding to $(\omega_X,\xi)=(-1, 3)$,  is ruled out at $\sim 95\%$ CL. However, the scenario of $\xi+3\omega_X < 0$ is supported at  $\sim 95\%$ CL, that corresponds to the non-standard cosmology with the energy transferring from dark energy to matter. The constraints on the parameters are more precision than those from \cite{Guo_et_al_2007,Chen_et_al_2010,Cao_et_al_2011}; however, the case with $\xi+3\omega_X >0$  is slightly preferred than the one with $\xi+3\omega_X <0$ in the these previous studies, that is different from the result of this work. 

(v) According to the Bayesian evidences calculated from the SNe+BAO+CMB sample, one can find out that the $\Lambda $CDM model is most supported; besides, the $\omega_0\omega_a$CDM model is more competitive than the $\xi$-index
model.

It's remarkable that the mock  GW siren sample can notably improve the precision on constraining $H_0$, and also be helpful in breaking the parameter degeneracies existing in the constraints from the SNe+BAO+CMB sample. In view of the above mentioned facts, it is promising that the future GW data will play a crucial role in estimating cosmological parameters.

\section*{Acknowledgments}
We would like to thank Minghui Du for supplying the mock catalog of BNS-SGRB GW events and also for some helpful suggestions on the use of the mock sample. This work has been supported by the National Natural Science Foundation of China
(Nos. 11988101, 12021003, 12033008, 11633001, 11920101003, 11703034, 11773032 and 11573031), the Strategic Priority Research
Program of the Chinese Academy of Sciences (No. XDB23000000), the Interdiscipline Research Funds
of Beijing Normal University, and the K. C. Wong Education Foundation.

\section*{Data Availability}
The data underlying this paper will be shared on reasonable request to the corresponding author.

\end{document}